\documentclass[superscriptaddress, aps, prd, reprint, nofootinbib]{revtex4-2}

\usepackage{graphicx}% Include figure files
\usepackage{amsmath,amssymb,mathtools,booktabs}
\usepackage{dcolumn}% Align table columns on decimal point
\usepackage{bm}% bold math
\usepackage{xcolor}
\usepackage{soul}
\usepackage{braket}
\usepackage{scalerel}

\usepackage[multi-part-units=single,group-digits=none]{siunitx}

\definecolor{Mathematica1}{rgb}{0.368417, 0.506779, 0.709798}
\definecolor{Mathematica2}{rgb}{0.880722, 0.611041, 0.142051}
\definecolor{Mathematica3}{rgb}{0.560181, 0.691569, 0.194885}
\definecolor{darkred}{rgb}{0.545,0,0}
\definecolor{dullblue}{rgb}{0,0.298,0.49}
\definecolor{blue3}{RGB}{31, 119, 180}
\usepackage[colorlinks=true
,urlcolor=dullblue
,anchorcolor=blue3
,citecolor=dullblue
,filecolor=blue3
,linkcolor=darkred
,menucolor=blue3
,pagecolor=blue3
,linktocpage=true
,pdfproducer=medialab
]{hyperref}

\newcommand{\dd}{\mathop{\mathrm{d}\!}{}}

\newcommand{\slab}[1]{{\textsc{#1}}}

\newcommand{\minus}{{\protect \scalebox {0.75}[1.0]{$-$}}}

\renewcommand{\vec}[1]{\boldsymbol{\mathbf{#1}}}

\newcommand{\ped}[1]{_{\mathrm{#1}}}
\renewcommand{\ap}[1]{^{\mathrm{#1}}}
\DeclarePairedDelimiter{\abs}{\lvert}{\rvert}

\def\beq{\begin{equation}}
\def\eeq{\end{equation}}

\begin{document}

\title{Probing dense environments around Sgr A* with S-star dynamics}

\author{Giovanni Maria Tomaselli}
\affiliation{School of Natural Sciences, Institute for Advanced Study, Princeton, NJ 08540, USA}

\author{Andrea Caputo}
\affiliation{Department of Theoretical Physics, CERN, 1217 Geneva, Switzerland}
\affiliation{Dipartimento di Fisica, ``Sapienza'' Università di Roma \& Sezione INFN Roma 1, 00185 Roma, Italy}
\affiliation{Department of Particle Physics and Astrophysics, Weizmann Institute of Science, Rehovot 7610001, Israel}

\begin{abstract}

The orbits of stars around Sgr~A*, the Milky Way’s supermassive black hole, provide a unique laboratory for testing its environment with unprecedented precision. In this work, we compute the apsidal precession induced by extended matter distributions through Lagrange’s equations and compare it with the measured precession of S2, reproducing and extending GRAVITY’s constraints. In particular, we push bounds on boson clouds to larger gravitational couplings $\alpha$ and to the second-fastest superradiant mode. We also show that environments with mass of order $1\%$ of Sgr~A* drive stellar orbits to decay by dynamical friction within a few Myr. The inner star cluster is, however, efficiently replenished, masking this effect observationally. We also prove that orbital resonances from boson clouds have no impact on relevant timescales. While S2 currently provides the cleanest dataset, our framework is readily applicable to other stars that we identify as particularly promising, whose orbits will be measured with increasing accuracy, opening up new opportunities to probe the environment of Sgr~A*.

\end{abstract}

\maketitle

\section{Introduction}

The strongest evidence that the compact radio source Sgr A* at the center of the Milky Way is associated to a supermassive black hole (BH) comes from the motion of the stars around it. Initial studies of the stellar velocity field in the central parsec of the Galaxy have eventually evolved into precise measurements of the acceleration and orbital parameters of dozens of stars. The currently most accurate data have been obtained with the GRAVITY interferometer at the European Southern Observatory’s (ESO) Very Large Telescope Interferometer (VLTI) \cite{2017A&A...602A..94G}, which have led to a measurement of the mass of Sgr A* equal to $M=\num{4.297\pm0.012e6}M_\odot$ \cite{GRAVITY:2021xju}.

The orbit of one of the stars, known as S2, has been measured so precisely that its general relativistic apsidal precession of about $\ang{;12;}$ per orbit has been detected with an extremely high degree of confidence \cite{GRAVITY:2020gka,GRAVITY:2021xju}. This measurement has then been used to place constraints on extended mass distributions around Sgr A* \cite{GRAVITY:2021xju,GRAVITY:2023cjt,GRAVITY:2023azi,GRAVITY:2024tth}, because any deviation from Newton's potential would affect the precession rate of S2.

In this work, we use the term ``environment'' to refer to any dark (i.e., non-luminous) extended mass distribution around Sgr A*. The existence of these environments, which can be much denser than the stellar cluster, is particularly interesting from the point of view of both astrophysics and fundamental physics. Astrophysical examples are given by a cluster of old stars and stellar remnants \cite{1972ApJ...178..371P,Frank:1976uy,1976ApJ...209..214B, BahcallWolf1976}, or a massive companion of Sgr A*. On the other hand, new ultralight bosons can create massive clouds through superradiance \cite{Arvanitaki:2010sy,Brito:2015oca}, while dark matter present in the halo can be compressed in a density ``spike'' around the BH \cite{Gondolo:1999ef,Ullio:2001fb,Gnedin:2003rj}. Current studies with S2 have tested various kinds of spherically symmetric density profiles, as well as the axially symmetric distributions expected in boson clouds.

In this paper, we adopt a perturbative approach to calculate the apsidal precession, based on Lagrange's equations~\cite{MurrayDermott1999}. This allows us to determine the environment-induced precession without the need to numerically integrate the stellar orbits. We use these results to revisit the constraints on the mass and density of environments around Sgr A*. Compared to Ref.~\cite{GRAVITY:2023cjt}, we improve the limits on the mass of a scalar cloud, probing values of the boson-BH gravitational coupling for which the cloud could form within a Hubble time.

While the apsidal precession has been extensively studied, other kinds of secular interactions between the environment and the stars have received comparatively little attention. In this paper we consider dynamical friction, showing that it can lead the stars with smallest pericenter distance to decay into Sgr A* on a timescale of a few $\si{Myr}$. We also consider orbital resonances induced by boson clouds, finding that they cannot be effectively excited as long as the star's primary energy loss mechanism is dynamical friction.

To check whether dynamical friction can be used to probe dense environments, we carefully compare it to the most important dynamical timescales of the star cluster. We then perform a simple simulation of the evolution of the cluster, studying how dynamical friction alters the distribution of orbital parameters of the stars. We find that the effect of dynamical friction is very subtle and probably unlikely to leave observational consequences, even when a good number of stars completely decay into the BH.

The paper is structured as follows: in Sec.~\ref{sec:environments} we review the various BH environments, the properties of the S-stars closest to Sgr A*, and the profile of the star cluster. In Sec.~\ref{sec:interactions} we calculate the environment-induced apsidal precession and use it to place constraints on its mass and density. We then compute the timescale required for dynamical friction to dissipate the star's orbital energies, and study the possible excitation of orbital resonances. In Sec.~\ref{sec:evolution} we simulate the cluster's evolution under the effect of dynamical friction. We conclude in Sec.~\ref{sec:conclusions}. Appendices~\ref{app:precession}, \ref{app:constraints-enclosed-mass}, \ref{sec:timescales}, and~\ref{app:pop-evol-no-srr} contain technical details on the calculation of precession, the environment's mass constraints, the stellar cluster timescales, and the effect of scalar resonant relaxation, respectively.

We use natural units $G=c=\hbar=1$ throughout.

\section{Dense environments and S-stars}
\label{sec:environments}

In this section, we delineate the astrophysical systems that are the subject of this paper. In Secs.~\ref{sec:superradiance} and \ref{sec:other-environments}, we describe the possible dark dense environments around Sgr A* considered in this work. Then, in Sec.~\ref{sec:s-stars}, we discuss the S-stars that are most suitable to probe such environments, and in Sec.~\ref{sec:cusp} we detail the density profile of the whole stellar cluster, which will be useful for some of our computations.

\subsection{Superradiant boson clouds}
\label{sec:superradiance}

One of the most studied and well motivated kinds of dark BH environment arises naturally in the presence of an ultralight boson field with mass $\mu$, such as an axion-like particle \cite{Wilczek:1977pj,Arvanitaki:2009fg}. In this paper, we focus on scalars, although much of the discussion can be extended to vector fields with little effort. When the BH spins rapidly enough, the so-called superradiant instability drives an exponential growth of certain scalar modes \cite{Starobinsky:1973aij,Arvanitaki:2010sy,Brito:2015oca, Detweiler:1980uk, East:2017ovw, Dolan:2007mj, cardoso2018constraining}. This instability eventually spins the BH down, saturates, and leaves a ``boson cloud'' with mass $M\ped{c}$ around the BH. The value of $M\ped{c}$ depends on the initial mass and spin of the BH, and it is at most $10\%M$ for a cloud that grows in isolation. This limit is increased by accretion onto the BH ~\cite{Brito:2014wla,Hui:2022sri}, and is reduced by the scalar self-interactions ~\cite{Baryakhtar:2020gao, Witte:2024drg, Hoof:2024quk}.

It is useful to define the gravitational coupling between the boson and the BH as $\alpha=\mu M$. For $\alpha\ll1$, the cloud is mostly nonrelativistic, and the scalar states can be described by hydrogenic wavefunctions $\ket{n\ell m}$, where $n$ is the principal quantum number and $\ell$, $m$ are the angular momentum and azimuthal numbers. In this paper we focus on the two fastest-growing states, $\ket{211}$ and $\ket{322}$. Their $e$-folding growth times can be approximated for $\alpha\ll1$ with Detweiler's approximation \cite{Detweiler:1980uk,Baumann:2019eav}, which gives
\begin{align}
T_{211}&\approx24M(1-2\alpha)^{-3}\alpha^{-9}\,,\\
T_{322}&\approx\frac{885735}{256}M(1-\alpha)^{-5}\alpha^{-13}\,.
\end{align}
Growing the cloud from a quantum fluctuation of the scalar field requires about 200 $e$-folds. For the mass of Sgr A*, this implies that the total growth time is shorter than $\SI{10}{Gyr}$ only when $0.042<\alpha<0.5$ for $\ket{211}$, and when $0.172<\alpha<1$ for $\ket{322}$.

The spatial profile of the cloud has a characteristic size
\beq
r\ped{c}=M/\alpha^2\,,
\eeq
with the density decreasing exponentially for large $r$. In particular, for $\alpha\ll1$, the densities of the $\ket{211}$ and $\ket{322}$ states are
\begin{align}
\label{eqn:density-cloud-211}
\rho_{211}(\vec r)&=\frac{M\ped{c}}{64\pi r\ped{c}^3}\biggl(\frac{r}{r\ped{c}}\biggr)^2e^{-r/r\ped{c}}\sin^2\theta\,,\\
\label{eqn:density-cloud-322}
\rho_{322}(\vec r)&=\frac{M\ped{c}}{26244\pi r\ped{c}^3}\biggl(\frac{r}{r\ped{c}}\biggr)^4e^{-2r/(3r\ped{c})}\sin^4\theta\,,
\end{align}
where $\theta$ is the polar angle with respect to the BH spin. Other bound states $\ket{n\ell m}$, even if not initially grown by superradiance, can be populated when a time-dependent tidal perturbation transfers part of the cloud from its initial state into other ones \cite{Baumann:2018vus}. Unbound states $\ket{k\ell m}$, where $k$ is the wavenumber, can also be similarly excited~\cite{Baumann:2021fkf}.

In this paper, we study how the orbits of S-stars are influenced by the density profiles \eqref{eqn:density-cloud-211} and \eqref{eqn:density-cloud-322}, as well as by the backreaction of their tidal perturbation to the cloud. Given the currently very large uncertainties on the orientation of the spin of Sgr A*, we assume for simplicity that the orbits of the S-stars lie in the equatorial plane. This is also generally the configuration where the cloud's dynamical effects are maximized. For a discussion that takes into account different values of the inclination, see Ref.~\cite{GRAVITY:2019tuf}.

\subsection{Other dark environments}
\label{sec:other-environments}

Supermassive BHs can also potentially harbor other kinds of environments. These can come in the form of a cluster of stellar BHs and other stellar remnants as considered in Ref.~\cite{GRAVITY:2024tth}, or more exotic environments like dark matter overdensities. A toy model that is often used in this context is the Plummer spherically symmetric density profile \cite{1911MNRAS..71..460P},
\beq
\rho\ped{pl}(r)=\frac{3M\ped{pl}}{4\pi r\ped{pl}^3}\biggl(1+\frac{r^2}{r\ped{pl}^2}\biggr)^{-5/2}\,,
\label{eqn:density-plummer}
\eeq
where $M\ped{pl}$ is the total mass of the environment and $r\ped{pl}$ is known as the scale radius.

Another well-motivated density distribution is a power-law one,
\beq
\rho_0(r)=\rho_0\biggl(\frac{r_0}r\biggr)^{\gamma_0}\,.
\label{eqn:density-spike}
\eeq
An example of dark environment with a power-law density profile are dark matter ``spikes'' \cite{Gondolo:1999ef,Ullio:2001fb,Gnedin:2003rj,Merritt:2002vj, Eda:2013gg, Vasiliev:2007vh,Eda:2014kra,Lacroix:2018zmg,Kavanagh:2020cfn, Bertone:2024wbn, Cole:2022ucw,Cheng:2024mgl}, which form when the supermassive BH grows adiabatically at the center of the halo. In this case, the power-law index is expected to be $2.25<\gamma_0<2.5$.

\subsection{S-stars with small pericenter distance}
\label{sec:s-stars}

The Sgr A* cluster, or S-cluster, contains several stars on a close orbit around the Milky Way's supermassive BH. The orbit of a given star with mass $M_\star$ can be described by its semi-major axis $a$, eccentricity $e$, and orbital period $T=2\pi\sqrt{a^3/M}$. The pericenter distance $r\ped{p}=a(1-e)$ plays an especially important role in our analysis. Indeed, a BH environment with characteristic size $r\ped{env}$ can hardly be probed by stars whose orbit is at all times at radii $r\gg r\ped{env}$, as the environment's mass $M\ped{env}$ would just be degenerate with that of Sgr A*. Nonspherical density profiles, such as Eqs.~\eqref{eqn:density-cloud-211} and \eqref{eqn:density-cloud-322}, generate a gravitational potential with quadrupole terms that are not degenerate with the BH's gravity; however, these higher-multipole effects decay very fast at large distances.

A star with pericenter distance $r\ped{p}$ can thus probe environments with size $r\ped{env}\gtrsim r\ped{p}$. Accurate tracking of stellar orbits has indeed been recently used to constrain extended mass distributions enclosed between the pericenter and apocenter of the tracked stars \cite{GRAVITY:2023cjt,GRAVITY:2023azi,GRAVITY:2024tth}. In the case of the best studied star, S2, this region corresponds to $2800<r/M<46000$ \cite{GRAVITY:2020gka}.

\begin{table}
\centering
\begin{tabular}{lccccc}
\toprule
Star & $a$\;[$\SI{e-3}{pc}$] & $e$ & $r\ped{p}$\;[$M$] & $T$\;[yr] & $M_\star$\;[$M_\odot$]\\
\midrule
S2 & $\num{5.016\pm0.005}$ & $\num{0.88444\pm0.00006}$ & $\num{2819\pm9}$ & \num{16.07\pm0.02} & 14\\
S29 & $\num{15.66\pm0.04}$ & $\num{0.96880\pm0.00009}$ & $\num{2376\pm11}$ & \num{88.6\pm0.3} & 10?\\
S38 & $\num{5.717\pm0.006}$ & $\num{0.81807\pm0.00022}$ & $\num{5058\pm16}$ & \num{19.55\pm0.03} & -\\
S55 & $\num{4.182\pm0.005}$ & $\num{0.72980\pm0.00018}$ & $\num{5496\pm17}$ & \num{12.23\pm0.02} & -\\
\color{gray}{S62} & \color{gray}\num{3.59\pm0.02} & \color{gray}\num{0.976\pm0.01} & \color{gray}\num{419\pm174} & \color{gray}\num{9.73\pm0.08} & \color{gray}6.1\\
\color{gray}S4711 & \color{gray}\num{3.00\pm0.06} & \color{gray}\num{0.768\pm0.03} & \color{gray}\num{3385\pm443} & \color{gray}\num{7.4\pm0.2} & \color{gray}2.2\\
\color{gray}S4714 & \color{gray}\num{4.079\pm0.012} & \color{gray}\num{0.985\pm0.011} & \color{gray}\num{298\pm218} & \color{gray}\num{11.78\pm0.05} & \color{gray}2.0\\
\bottomrule
\end{tabular}
\caption{Semi-major axis $a$, eccentricity $e$, pericenter distance $r\ped{p}$, orbital period $T$ and mass $M_\star$ of selected stars around Sgr A*. The data for S2, S29, S38, and S55 are from Ref.~\cite{GRAVITY:2024tth}, while the data for S62, S4711, and S4714 are from Ref.~\cite{2020ApJ...899...50P}. We report the latter three stars in gray shade, because subsequent analyses failed to reproduce the measurements of S62 \cite{GRAVITY:2020qsl,GRAVITY:2021xxx}. To report all parameters with consistent units across all stars, we converted and/or determined the values of $a$ and $T$ using the distance to Galactic Center $R_0=\SI{8.2759\pm0.0086}{kpc}$ and the mass of Sgr A* $M=\num{4.297\pm0.012e6}M_\odot$. We ignored systematic uncertainties, which are of the order of $\SI{0.03}{kpc}$ for $r\ped{s}$ and $\num{0.04e6}M_\odot$ for $M$ \cite{GRAVITY:2021xju}. The mass of S29 has not been well determined in the literature, but we assume a value of $10M_\odot$ for illustrative purposes.}
\label{tab:stars}
\end{table}

We list in Table~\ref{tab:stars} the orbital parameters of the S-stars with smallest pericenter distance. S29 is the star with the closest pericenter, followed by S2, which has, however, a significantly shorter period. Whether stars with even closer pericenter exist is currently unclear. Two candidates with $r\ped{p}\sim400M$, namely S62 and S4714, were reported in \cite{2020ApJ...889...61P,2020ApJ...899...50P}. However, subsequent observations suggested that the orbit of S62 was miscalculated, due to the star passing close to Sgr A* in projection, while remaining at a large physical distance \cite{GRAVITY:2020qsl,GRAVITY:2021xxx}. We thus report with a gray shade in Table~\ref{tab:stars} the parameters of the stars discovered in Refs.~\cite{2020ApJ...889...61P,2020ApJ...899...50P}.

The possible existence of stars with pericenter closer than that of S29 and S2 is particularly relevant to probe superradiant clouds. Indeed, in the case of S2, we have $r\ped{c}>r\ped{p}$ only for $\alpha\lesssim0.03$. This value is outside the window within which the cloud can grow within a Hubble time, as we determined in Sec.~\ref{sec:superradiance}. Within the physically motivated parameter space, stars such as S29 and S2 can therefore only skim the cloud's outskirts. For this reason, the constraining power of these stars mostly falls outside the relevant range of values of $\alpha$. Stars such as S62 and S4714 would instead be ideally suited to probe $\alpha\sim\mathcal O(0.1)$.

\subsection{Stellar cusp profile}
\label{sec:cusp}

Apart from the orbital parameters of individual stars, it is also useful to know the density profile of the stellar cluster as a whole~\cite{Peebles1972, Frank:1976uy, BahcallWolf1976}. From observations at distances $r\lesssim\SI{0.5}{pc}$ from Sgr A*, and assuming a constant mass-to-light ratio, the authors of Ref.~\cite{Schodel:2017vjf} determined that the density profile of the stellar cluster is well approximated by
\beq
\rho(r)=\rho\ped{s}\biggl(\frac{r\ped{s}}r\biggr)^{\gamma\ped{s}}\,,
\label{eqn:density-stellar-cusp}
\eeq
with $\rho\ped{s}=\num{2.6e7}\,M_\odot/\text{pc}^3$, $r\ped{s}=\SI{0.01}{pc}$, and $\gamma\ped{s}=1.13$. The corresponding stellar mass enclosed within a radius $r$ is $M\ped{s}(r)=M(r/r_h)^{3-\gamma\ped{s}}$, where
\beq
r_h=r\ped{s}\biggl(\frac{(3-\gamma\ped{s})M}{4\pi\rho\ped{s}r\ped{s}^3}\biggr)^{1/(3-\gamma\ped{s})}=\SI{2.23}{pc}
\eeq
is the radius of influence of Sgr A*.

\section{Star-environment interaction}
\label{sec:interactions}

\subsection{Apsidal precession}
\label{sec:apsidal-precession}

An extended mass distribution around Sgr A* generates a gravitational potential $V(r)$ that deviates from the point-mass $r^{-1}$ law, causing the pericenter of S-stars to precess. The secular evolution of the longitude of pericenter $\varpi$ can be determined using Lagrange's equation in perturbation theory \cite{MurrayDermott1999} as
\beq
\dot\varpi=-\frac{\sqrt{1-e^2}}{\sqrt{Ma}\,e}\frac{\partial\langle V\rangle}{\partial e}\,,
\label{eqn:dotvarpi}
\eeq
where the time average of $V(r)$ along the orbit is
\beq
\begin{split}
\braket{V}&=\frac1T\int_0^TV(r(t))\dd t\\
&=\frac1{2\pi}\int_0^{2\pi}(1-e\cos E)V(r(E))\dd E\,,
\end{split}
\eeq
where in the second equality we introduced the eccentric anomaly $E$, related to $r$ via $r(E)=a(1-e\cos E)$. Equation~\eqref{eqn:dotvarpi} holds for spherically symmetric potentials and for equatorial orbits in axisymmetric potentials, while more general configurations require taking into account the orbital inclination.

Given the density $\rho(r)$ of a spherically symmetric environment, the gravitational potential is
\beq
V(r)=\int_0^r\frac{\dd r'}{r'^2}\int_0^{r'}4\pi r''^2\rho(r'')\dd r''\,,
\eeq
which we can plug in Eq.~\eqref{eqn:dotvarpi} to determine $\dot\varpi$. For a power-law profile such as Eq.~\eqref{eqn:density-spike}, we can express the apsidal precession rate in terms of Gauss' hypergeometric function as
\beq
\dot\varpi=-T\sqrt{1-e^2}\rho_0\biggl(\frac{r_0}a\biggr)^{\gamma_0}{}_2F_1\biggl(\frac{\gamma_0-1}{2}, \frac{\gamma_0}{2}; 2; e^2\biggr)\,.
\label{eqn:varpi-power-law}
\eeq
The overall minus sign shows that the apsidal precession is retrograde, as in any spherical environment \cite{Tremaine:2004yu}. For the Plummer profile, given in Eq.~\eqref{eqn:density-plummer}, the time average integral cannot be solved in closed form, so we determine $\dot\varpi$ numerically. The apsidal precession induced by all spherical environments is caused by the fact that the environment's enclosed mass varies as the star moves along its eccentric orbit.

\begin{figure*}
\centering
\includegraphics[]{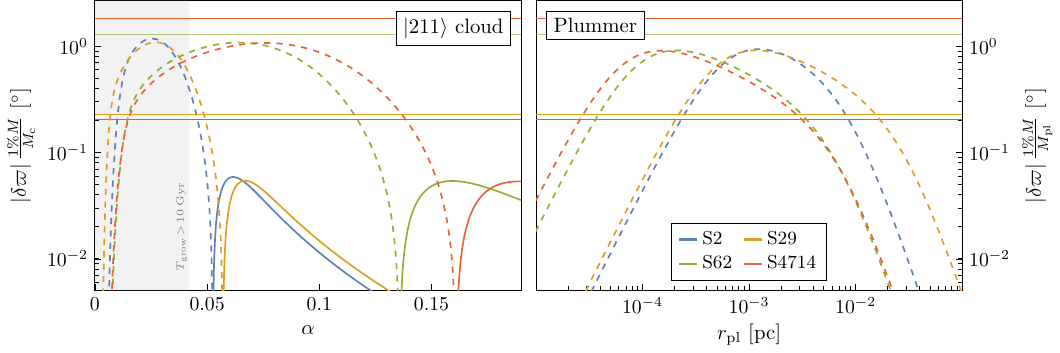}
\caption{Environment-induced apsidal precession per orbit $\delta\varpi=\dot\varpi T$ for four representative S-stars, for the case of a scalar cloud (\emph{left panel}) and a Plummer profile (\emph{right panel}), as a function of $\alpha$ and $r\ped{pl}$ respectively. Solid (dashed) lines indicate $\delta\varpi>0$ ($\delta\varpi<0$). The thin horizontal lines show the Schwarzschild precession \eqref{eqn:gr-precession}. The gray band (\emph{left panel}) indicates the values of $\alpha$ for which the cloud's growth takes longer than $\SI{10}{Gyr}$.}
\label{fig:precession-per-orbit}
\end{figure*}

The density of a boson cloud, given in Eqs.~\eqref{eqn:density-cloud-211} and~\eqref{eqn:density-cloud-322}, is axially (not spherically) symmetric. Its gravitational potential is of the form
\beq
\begin{split}
V(\vec r)={}&V_0(r)+(3\cos^2\theta-1)V_2(r)\\
&+(35\cos^4\theta-30\cos^2\theta+3)V_4(r)\,,
\end{split}
\label{eqn:potential-cloud-short}
\eeq
where $V_4(r)$ vanishes for $\ket{211}$. The second term in Eq.~\eqref{eqn:potential-cloud-short} gives rise to a quadrupole field, which decays as $V_2\sim r^{-3}$, while the third term is a hexadecapole, $V_4\sim r^{-5}$. The monopole term $V_0$ gives rise to apsidal precession through the variation of the cloud's mass enclosed within a radius $r$; this contribution to $\dot\varpi$ decays exponentially with $a$. On the other hand, the quadrupole and hexadecapole fields decay as power laws, and thus cause a precession even of the stellar orbits that do not plunge inside the densest regions of the cloud. More specifically, in the limit of large semi-major axis, the precession rate decays as $\dot\varpi\sim a^{-7/2}$, rather than exponentially (for the vector case, see Ref.~\cite{Cao:2024wby}). Furthermore, these terms can cause prograde precession, which is never observed for spherical profiles. In Appendix~\ref{app:precession} we write the explicit expression of Eq.~\eqref{eqn:potential-cloud-short}, and also give further details on the calculation of $\dot\varpi$ in spherical environments. Note that our perturbative approach is equivalent to that presented in Ref.~\cite{GRAVITY:2019tuf}.\footnote{We have thoroughly checked that our analytical expressions agree with those of Ref.~\cite{GRAVITY:2019tuf}, although they do not provide a compact formula for the apsidal precession, like our Eq.~\eqref{eqn:dotvarpi-cloud}, nor use their results to place bounds on the environment's mass. Nevertheless, we note that, for small-$\alpha$, their Fig.~3 implies an apsidal precession much larger than that shown in our Fig.~\ref{fig:precession-per-orbit}. We do not know the origin of this discrepancy.}

We show in Fig.~\ref{fig:precession-per-orbit} the apsidal precession per orbit, $\delta\varpi=\dot\varpi T$, of S2, S29, S62, and S4714, for the cases of a $\ket{211}$ cloud and a Plummer profile. We note that the peak magnitude of $\delta\varpi$ is very similar across all stars. However, different stars are able to probe environments with different characteristic scales, based on the relation $r\ped{p}\gtrsim r\ped{env}$ as argued in Sec.~\ref{sec:s-stars}. While the monopole-induced retrograde precession dominates for $a\sim r\ped{c}$, for $a/r\ped{c}\gg 1$ (large $\alpha$) the quadrupole field takes over, and the precession becomes prograde.

The orbits of S-stars precess even in absence of any dark environment. The general-relativistic (or ``Schwarzschild'') precession rate is
\beq
\frac{2\pi}{\dot\varpi_\slab{gr}}=\frac{r\ped{p}(1+e)T}{3M}=\begin{cases}
\SI{28.4}{kyr} & \text{S2},\\
\SI{138}{kyr} & \text{S29},\\
\SI{2.68}{kyr} & \text{S62},\\
\SI{2.33}{kyr} & \text{S4714}.\\
\end{cases}
\label{eqn:gr-precession}
\eeq
The stellar cluster itself, whose density is given in Eq.~\eqref{eqn:density-stellar-cusp}, also contributes to the precession rate with an expression analogous to Eq.~\eqref{eqn:varpi-power-law} (replacing $\rho_0$, $r_0$ and $\gamma_0$ with $\rho\ped{s}$, $r\ped{s}$ and $\gamma\ped{s}$), which gives
\beq
\frac{2\pi}{\dot\varpi_0}=\begin{cases}
-\SI{3.23}{Myr} & \text{S2},\\
-\SI{3.96}{Myr} & \text{S29},\\
-\SI{7.77}{Myr} & \text{S62},\\
-\SI{9.36}{Myr} & \text{S4714}.\\
\end{cases}
\eeq
Because $\abs{\dot\varpi_0}\ll\dot\varpi_\slab{gr}$ for all the stars considered, we can safely neglect the cluster-induced precession.

\begin{figure*}
\centering
\includegraphics[]{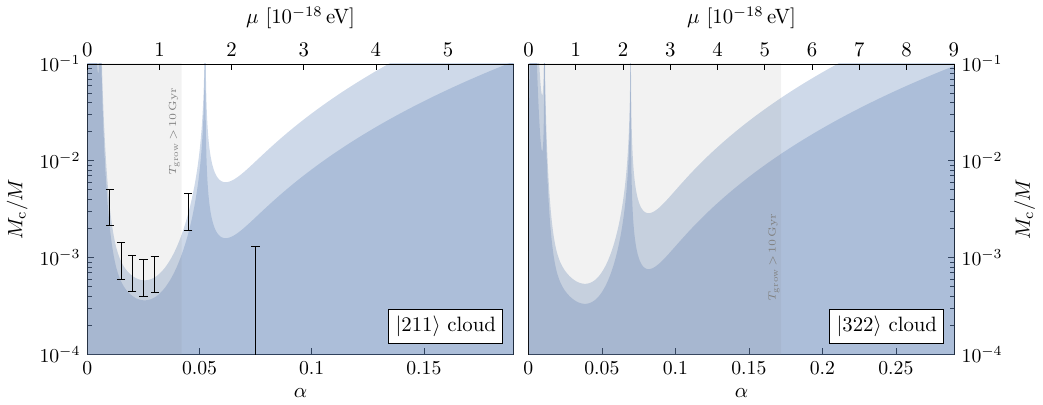}
\caption{1-$\sigma$ (dark blue) and 2-$\sigma$ (light blue) constraints on the mass of a boson cloud around Sgr A* in the $\ket{211}$ (\emph{left panel}) and $\ket{322}$ (\emph{right panel}) states, as a function of $\alpha=\mu M$. The contours are calculated based on the apsidal precession of S2, as measured in Ref.~\cite{GRAVITY:2024tth}, and assume equatorial orbits. Constraints from the GRAVITY Collaboration \cite{GRAVITY:2023cjt}, based on an MCMC analysis, are shown as black error bars (\emph{left panel})---these seem to indicate a 2-$\sigma$ preference for a nonzero cloud's mass for $0.01<\alpha<0.05$, which was however dismissed in Ref.~\cite{GRAVITY:2023cjt} as the Bayesian evidence was strong but not decisive. The gray band indicates the values of $\alpha$ for which the cloud's growth takes longer than $\SI{10}{Gyr}$.}
\label{fig:precession-bounds-cloud}
\end{figure*}

\begin{figure*}
\centering
\includegraphics[]{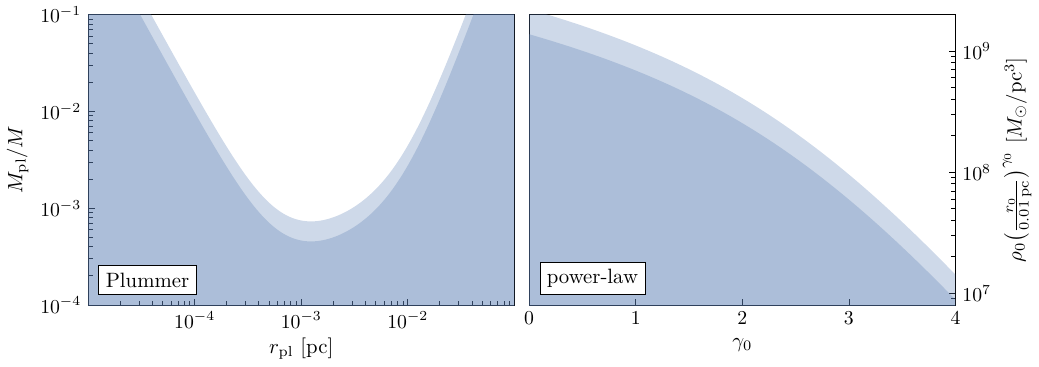}
\caption{1-$\sigma$ (dark blue) and 2-$\sigma$ (light blue) constraints on the environment mass from the apsidal precession of S2, for the case of Plummer (\emph{left panel}) and power-law (\emph{right panel}) density profiles. We constrain the total mass $M\ped{pl}$ of the Plummer profile as a function of the Plummer scale radius $r\ped{pl}$, as defined in Eq.~\eqref{eqn:density-plummer}, and the density $\rho_0$ at $\SI{0.01}{pc}$ as a function of the power-law index $\gamma_0$, as defined in Eq.~\eqref{eqn:density-spike}. Note that, even though we show here the constrain on the parameter $M\ped{pl}$ of Eq.~\eqref{eqn:density-plummer}, the motion of S2 is only sensitive to the mass enclosed between its pericenter and apocenter, and not to the total environment's mass. We show the bound on such enclosed mass in Appendix~\ref{app:constraints-enclosed-mass}.}
\label{fig:precession-bounds-plummer-spike}
\end{figure*}

The apsidal precession of S2 has been observed for the first time in Ref.~\cite{GRAVITY:2020gka}. The latest measurements \cite{GRAVITY:2024tth} show that the periapsis of S2 precesses at a rate equal to $f\dot\varpi_\slab{gr}$, where $f=0.918\pm0.128$. Given that the total apsidal precession rate is $\dot\varpi\ped{tot}=\dot\varpi_\slab{gr}+\dot\varpi$, this measurement can directly used to constrain environments around Sgr A*. Because $\dot\varpi$ is proportional to the mass of the environment, say $M\ped{c}$ or $M\ped{pl}$, the observational upper bound on these quantities scales inversely with the astrometric uncertainty on $\delta\varpi$. In particular, we can calculate the $n$-$\sigma$ constraint by imposing $\dot\varpi\ped{tot}=(f\pm n\sigma_f)\dot\varpi_\slab{gr}$, and then solving for the mass or the density of the environment.

This orbit-averaged approach is much simpler but less informative than a full orbital fit, which was performed, e.g., in Refs.~\cite{GRAVITY:2023cjt} and~\cite{GRAVITY:2024tth}. However, as we now show, these two approaches actually give almost identical constraints.

We show the constraints obtained this way in Figs.~\ref{fig:precession-bounds-cloud} and~\ref{fig:precession-bounds-plummer-spike}. Since $f$ is compatible with 1, all our constraints are compatible with a null detection of an environment at 1-$\sigma$.

For the case of a $\ket{211}$ boson cloud, we find that the strongest constraints lie in the unphysical region $\alpha<0.042$ (where $T\ped{grow}=200 \, T_{211}>\SI{10}{Gyr}$). We still find, however, nontrivial constraints at larger values of $\alpha$, which for example limit $M\ped{c}$ to be less than $0.01M$ at 1-$\sigma$ for $\alpha<0.1$. The Markov chain Monte Carlo (MCMC) analysis of the GRAVITY Collaboration \cite{GRAVITY:2023cjt} seems to suggest instead a preference for a nonzero cloud's mass at about 2-$\sigma$ for $0.01<\alpha<0.05$, shown as black error bars in Fig.~\ref{fig:precession-bounds-cloud} (\emph{left panel}). This was dismissed in Ref.~\cite{GRAVITY:2023cjt} due to the strong but not decisive Bayesian evidence. We do not know why the results of Ref.~\cite{GRAVITY:2023cjt} are not compatible with zero, but we note that their central values closely track our 2-$\sigma$ bound as $\alpha$ is varied between 0 and $0.05$.

The constraints for a $\ket{322}$ cloud largely lie in the unphysical region $\alpha<0.172$ (where $T\ped{grow}=200T_{322}>\SI{10}{Gyr}$). Nevertheless, in this case too we find nontrivial constraints at larger values of $\alpha$, for example limiting $M\ped{c}$ to be less than $0.02M$ at $\alpha<0.2$.

The constraints on the total mass of an environment with a Plummer profile are strongest when $r\ped{pl}\sim\SI{e-3}{pc}\sim2r\ped{p}$, where we find $M\ped{pl}\lesssim10^{-3}M$. The density of a power-law profile is constrained to be less than about 2 orders of magnitude denser than the stellar cusp, which requires $\rho\lesssim\num{e9}\,M_\odot/\si{pc^3}$ at $r=\SI{0.01}{pc}$ (roughly the apocenter distance of S2) for $0<\gamma_0<2$. Note that both these density profiles were studied in Ref.~\cite{GRAVITY:2024tth}, but their constraints focus on the mass enclosed between pericenter and apocenter of S2, and are thus not immediately comparable with Fig.~\ref{fig:precession-bounds-plummer-spike}. To allow for a more direct comparison, in Appendix~\ref{app:constraints-enclosed-mass} we translate our bounds into limits on such enclosed mass. The constraints shown in our Fig.~\ref{fig:precession-bounds-plummer-spike-enclosed} are almost identical to those in Fig.~B.3 of Ref.~\cite{GRAVITY:2024tth}, demonstrating that our simple, semi-analytical, orbit-averaged approach delivers constraints almost as powerful as a full orbital fit, at least given the current astrometric data.

\subsection{Dynamical friction}
\label{sec:df}

When a point mass $M_\star$ moves with velocity $v$ through a medium with density $\rho$, it experiences a drag force known as dynamical friction (DF), losing energy at a rate \cite{Chandrasekhar:1943ys}
\beq
P_\slab{df}=\frac{4\pi M_\star^2\rho}{v}\log\Lambda\,,
\label{eqn:Pdf}
\eeq
where the ``Coulomb logarithm'' $\log\Lambda$ is a factor of $\mathcal O(1-10)$ that depends on the geometry of the medium.\footnote{In Eq.~\eqref{eqn:Pdf}, we neglected extra factors that depend on the nature of the medium. These are not relevant for our estimate, and we will perform later a more accurate computation in the case of a boson cloud.} We study here for the first time the effect of DF from a dark environment around Sgr A* on the orbits of S-stars. Differently from the apsidal precession studied in Sec.~\ref{sec:apsidal-precession}, DF is a dissipative, rather than conservative, effect. As such, it leads to a slow decay of the stellar orbits.

Let us estimate the secular effect of DF on a star with $r\ped{p}\approx r\ped{env}$. On very eccentric orbits, like all the ones in Table~\ref{tab:stars}, the star only moves through the densest part of the environment when it is close to pericenter. Here, the velocity of the star is $v\approx\sqrt{2M/r\ped{p}}$ and the density is $\rho\approx M\ped{env}/\bigl(\frac43\pi r\ped{env}^3\bigr)$, where $M\ped{env}$ is the mass of the environment, which the star traverses in a time $t\ped{cross}\approx2r\ped{p}/v$. We then define the characteristic DF timescale $T_\slab{df}^{(0)}$ as the time required for DF to dissipate the orbital energy $E_\star=-M_\star M/(2a)$, as follows:
\beq
T_\slab{df}^{(0)}\equiv T\frac{\abs{E_\star}}{P_\slab{df}t\ped{cross}}=\frac{1-e}6\frac{M}{M\ped{env}}\frac{T}q\,,
\label{eqn:Tdf0}
\eeq
where $q=M_\star/M$ is the star-to-BH mass ratio and we neglected the Coulomb logarithm. The values of $T_\slab{df}^{(0)}$ for the four stars with the closest pericenter are
\beq
T_\slab{df}^{(0)}=\frac{0.01M}{M\ped{env}}\times\begin{cases}
\SI{9.5}{Myr} & \text{S2},\\
\SI{19.8}{Myr} & \text{S29},\\
\SI{2.7}{Myr} & \text{S62},\\
\SI{6.3}{Myr} & \text{S4714}.\\
\end{cases}
\eeq
The actual energy lost per orbit $E\ped{lost}$ will differ from $P_\slab{df}t\ped{cross}$ by $\mathcal O(1)$ factors, depending on the details of the environment. When $E\ped{lost}$ is known, we can calculate more accurately the DF timescale as
\beq
T_\slab{df}\equiv T\frac{\abs{E_\star}}{E\ped{lost}}\,.
\label{eqn:Tdf}
\eeq
As long as $E\ped{lost}$ only depends on $r\ped{p}$ (rather than $a$ and $e$ separately), $T_\slab{df}$ can be written as $T_\slab{df}^{(0)}$ times a function of $r\ped{p}/r\ped{env}$.

Because DF is only effective close to pericenter, we may approximate the orbital evolution by requiring that $r\ped{p}$ remains fixed, while $a$ and $T$ decay. The energy balance equation
\beq
\dot E_\star=-\frac{E\ped{lost}}T\,
\eeq
can then be solved as
\beq
T(t)=T(0)\biggl(1-\frac{t}{2T_\slab{df}(0)}\biggr)^3\,,
\label{eqn:T(t)_DF}
\eeq
where $T(0)$ is the initial orbital period, $T(t)$ is the orbital period after a time $t$, and $T_\slab{df}(0)$ is the initial value of $T_\slab{df}$.

\begin{figure}
\centering
\includegraphics[]{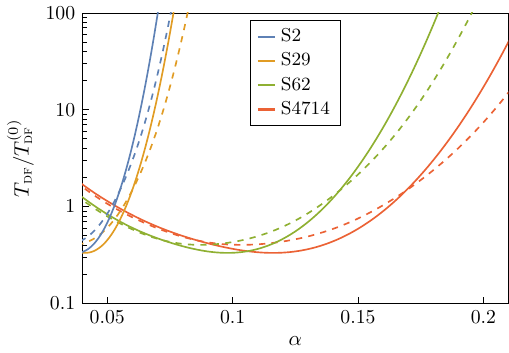}
\caption{Dynamical friction timescale $T_\slab{df}$, as defined in Eq.~\eqref{eqn:Tdf}, from a superradiant cloud in the $\ket{211}$ state on a star orbiting Sgr A*. We show the result normalized by the characteristic DF timescale $T_\slab{df}^{(0)}$, defined in Eq.~\eqref{eqn:Tdf0}, with $M\ped{env}=M\ped{c}$. Solid (dashed) lines indicate prograde (retrograde) orbits with respect to the cloud's spin.}
\label{fig:T_DF}
\end{figure}

An environment where $E\ped{lost}$ can be calculated exactly are boson clouds, where DF is also known as \emph{ionization} \cite{Baumann:2021fkf,Baumann:2022pkl}. Given the extremely high eccentricities of the stars in Table~\ref{tab:stars}, we use for simplicity the result for $E\ped{lost}$ on a parabolic orbit, as calculated in Ref.~\cite{Tomaselli:2023ysb}, and focus on the case of a $\ket{211}$ state. The energy lost can be written as
\beq
\begin{split}
E\ped{lost}={}&\frac{M\ped{c}}\mu\sum_{n\ell m}(E_{n\ell m}-E_{211})\abs{c_{n\ell m}}^2\\
&+\frac{M\ped{c}}\mu\int\frac{\dd k}{2\pi}\sum_{\ell m}(E_{k\ell m}-E_{211})\abs{c_{k\ell m}}^2\,,
\end{split}
\label{eqn:Elost-cloud}
\eeq
where $E_{n\ell m}$ is the energy of the state $\ket{n\ell m}$ and
\beq
c_{n\ell m}=-i\int_{-\infty}^\infty\dd t\braket{n\ell m|V_\star|211}e^{i(E_{n\ell m}-E_{211})t}\,,
\label{eqn:cnlm}
\eeq
and an identical formula holds for unbound states $\ket{k\ell m}$. The potential $V_\star$ is the gravitational perturbation induced by the star on the cloud, calculated in this case on a parabolic orbit with pericenter $r\ped{p}$. We refer to Ref.~\cite{Tomaselli:2023ysb} for further details about the calculation of the integral \eqref{eqn:cnlm}. The second term in Eq.~\eqref{eqn:Elost-cloud} is always positive, because unbound states have positive energies $E_{k\ell m}>0$, while $E_{211}<0$. The first term, instead, does not have a definite sign, due to the existence of some states ($\ket{100}$, $\ket{21\,\minus1}$ and $\ket{210}$) with energy smaller than that of $\ket{211}$. When $r\ped{p}\sim r\ped{c}$, the second term in Eq.~\eqref{eqn:Elost-cloud} always dominates, so that we do not have to worry about $E\ped{lost}$ potentially being negative. We show in Fig.~\ref{fig:T_DF} the ratio $T_\slab{df}/T_\slab{df}^{(0)}$ for the four stars with closest pericenter. As expected, this ratio is of $\mathcal O(1)$ as long as the star passes through the densest regions of the cloud ($r\ped{p}\lesssim r\ped{c}$), while $T_\slab{df}$ increases exponentially for large $r\ped{p}$.

\begin{figure*}
\centering
\includegraphics[]{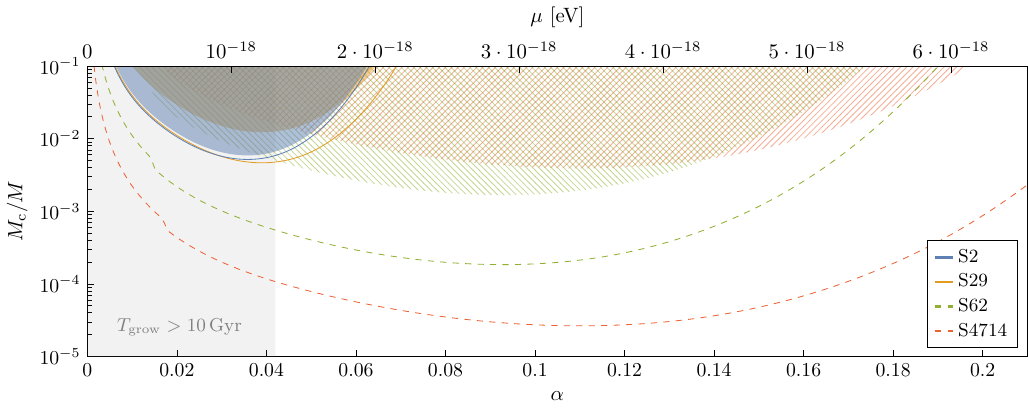}
\caption{Values of the mass $M\ped{c}$ of a boson cloud for which the orbits of S-stars experience significant decay due to dynamical friction, as a function of $\alpha$. The shaded areas indicate where the decay time is shorter than the age of the Galactic Center disc(s), $T_\slab{df}<\SI{6}{Myr}$ \cite{Paumard:2006im}. The thin lines indicate where the decay time becomes comparable to the stars' expected lifetimes, $T_\slab{df}=T\ped{life}/2$, assuming Eq.~\eqref{eq:MSlife} and the masses $M_\star$ reported in Table~\ref{tab:stars}. The contours and the lines for S62 and S4714 are dashed because their parameters as measured in Ref.~\cite{2020ApJ...899...50P} could not be confirmed by later studies \cite{GRAVITY:2020qsl,GRAVITY:2021xxx}. For simplicity, in this plot we use the average of $E\ped{lost}$ on prograde and retrograde orbits; the difference between the two cases is marginal, as shown in Fig.~\ref{fig:T_DF}. The gray band indicates the values of $\alpha$ for which the cloud's growth takes longer than $\SI{10}{Gyr}$.}% Constraints from tracking stellar orbits \cite{GRAVITY:2023cjt} are shown as black error bars---these seem to indicate the detection of a nonzero cloud's mass for $0.01<\alpha<0.05$, which was however dismissed in \cite{GRAVITY:2023cjt} as the Bayesian evidence was strong but not decisive.}
\label{fig:df-bounds}
\end{figure*}

As we demonstrate in Appendix~\ref{sec:timescales}, the DF timescale $T_\slab{df}$ can be comparable, or even shorter, than other typical timescales of the star cluster, such as the stellar lifetime, resonant relaxations and physical collisions. We show in Fig.~\ref{fig:df-bounds} the values of $M\ped{c}$ such that the orbits of S-stars would completely decay into Sgr A* in a time shorter than the age of the Galactic Center disc (approximately $\SI{6}{Myr}$ \cite{Paumard:2006im}), and in a time shorter than the expected lifetime of the stars themselves. As noted for the apsidal precession in Sec.~\ref{sec:apsidal-precession}, we see that stars with smaller pericenter are able to probe larger values of $\alpha$.

It is tempting to interpret the limits in Fig.~\ref{fig:df-bounds} as observational bounds on $M\ped{c}$, invoking a fine-tuning argument. This can be expressed as saying that it seems unlikely that we happen to observe a star whose orbit is about to decay in a short time, compared to the age of the star itself. This argument would only rely on approximate knowledge of the orbital parameters, rather than on high-precision astrometric data such as those needed to measure the apsidal precession discussed in Sec.~\ref{sec:apsidal-precession}. One should, however, keep in mind that DF would make stars on eccentric orbit continuously ``flow'' from large to small semi-major axes, thus replenishing the region where the decay time is short. It is therefore necessary to study the evolution of the whole star cluster to determine the expected impact of DF on the distribution of stellar orbits. We perform this task in Sec.~\ref{sec:evolution}.

\subsection{Star-cloud orbital resonances}

There is another possible kind of star-environment interaction which is specific to boson clouds: binary perturbers can give rise to orbital resonances \cite{Baumann:2018vus,Berti:2019wnn,Baumann:2019ztm,Takahashi:2021yhy,Takahashi:2023flk,Tomaselli:2024bdd,Takahashi:2024fyq,Tomaselli:2025jfo}. These occur when the energy difference between two states of the cloud equals an integer multiple of the binary frequency $\Omega=\sqrt{M/a^3}$,
\beq
E_{n'\ell'm'}-E_{n\ell m}=g\Omega\,.
\label{eqn:Omega-res}
\eeq
So-called Bohr resonances ($n'\ne n$) require the semi-major axis to be comparable with the size of the cloud, $a\sim r\ped{c}$. For all the stars listed in Table~\ref{tab:stars}, this implies $\alpha\lesssim0.01$, a value far too small for the cloud to grow within a Hubble time. On the other hand, the orbital periods corresponding to hyperfine resonances ($n'=n$, $\ell'=\ell$),
\begin{align}
\label{eqn:Thf211}
T\ped{hf}\ap{211}&=\SI{1.0}{yr}\,\biggl(\frac{0.2}{\alpha}\biggr)^7\frac{1+4\alpha^2}{1-m'}(-g)\,,\\
\label{eqn:Thf322}
T\ped{hf}\ap{322}&=\SI{33}{yr}\,\biggl(\frac{0.2}{\alpha}\biggr)^7\frac{1+\alpha^2}{2-m'}(-g)\,,
\end{align}
and fine resonances ($n'=n$, $\ell'\ne\ell$),
\beq
T\ped{fine}\ap{322}=\SI{0.67}{yr}\,\biggl(\frac{0.2}{\alpha}\biggr)^5(-g)\,,
\label{eqn:Tfine322}
\eeq
happen to lie in the range spanned by S-stars for acceptable values of $\alpha$.\footnote{The only fine resonance for the $\ket{211}$ state is extremely weak due to the large decay width of $\ket{200}$ \cite{Tomaselli:2024bdd}, so we do not list it here.}

Fine and hyperfine resonances are known to give rise to \emph{floating orbits}, wherein the frequency of an otherwise inspiralling object remains approximately constant. Furthermore, orbital parameters such as inclination and eccentricity evolve during the resonance, potentially approaching fixed points with nonzero eccentricity \cite{Boskovic:2024fga,Tomaselli:2024dbw}. This suggests the intriguing possibility of probing boson clouds through resonance-induced clustering of S-stars' orbital frequencies, inclinations and eccentricities.

Floating orbits occur when the orbital energy lost by the star (through any mechanism) is balanced by the energy gained from the cloud's state transition. At the resonant frequency \eqref{eqn:Omega-res}, the binary gains energy from the cloud at a rate \cite{Tomaselli:2025jfo}
\beq
P\ped{res}=\frac{M\ped{c}}{\mu}\frac{2(E_{n\ell m}-E_{n'\ell' m'})\abs{\braket{n'\ell' m'|V_\star|n\ell m}}^2}{\Gamma_{n'\ell' m'}}\,,
\eeq
where $\Gamma_{n'\ell'm'}$ is the decay rate of the state $\ket{n'\ell' m'}$ into the BH. In Sec.~\ref{sec:df} we showed that the DF from the cloud itself can be an efficient energy loss mechanism. As we show in Appendix~\ref{sec:timescales}, DF is indeed faster than most other timescales of the star cluster.

\begin{figure}
\centering
\includegraphics[]{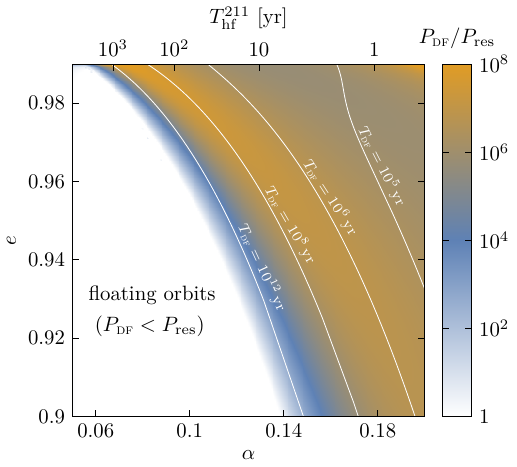}
\caption{Color map showing the ratio between dynamical friction power $P_\slab{df}$ and resonance power $P\ped{res}$ for a $\ket{211}$ cloud, at the resonant frequency \eqref{eqn:Thf211}, with $m'=g=-1$, as a function of $\alpha$ and $e$. The white area is where $P_\slab{df}<P\ped{res}$, i.e., floating orbits occur. The white contours show values of $T_\slab{df}$, calculated at the resonant frequency and assuming $M_\star=10M_\odot$ and $M_c=0.01M$. It is clear that floating orbits can only occur when $T_\slab{df}$ is extremely large. The same conclusion applies for other values of $m$ and $g$, since they require lower frequencies and thus even longer $T_\slab{df}$.}
\label{fig:Pdf_Pres}
\end{figure}

Following the method described in Ref.~\cite{Tomaselli:2025jfo}, we therefore compute the ratio $P_\slab{df}/P\ped{res}$. When this ratio is smaller than unity at the resonance frequency, the star enters a floating orbit, as it gains more energy than it loses (and will evolve towards a frequency fixed point, slightly above the resonant frequency, where the two contributions cancel each other and net energy change vanishes). Otherwise, the star always loses more energy than it gains, and the orbit keeps shrinking without any floating orbit occurring. Note that both $P_\slab{df}$ and $P\ped{res}$ depend linearly on the cloud's mass $M\ped{c}$, therefore their ratio is independent of it.

We show $P_\slab{df}/P\ped{res}$ in Fig.~\ref{fig:Pdf_Pres} as a function of $\alpha$ and $e$. The region where floating orbits occur ($P_\slab{df}<P\ped{res}$) turns out to require such a small $P_\slab{df}$ that $T_\slab{df}$ becomes longer than the age of the Universe, as shown with white contours in  Fig.~\ref{fig:Pdf_Pres}. We conclude that DF from the cloud itself can never effectively excite (hyper)hyperfine resonances. Should some other energy loss mechanism act on the stars, it could reopen the possibility of looking for resonances in the orbits of S-stars, given they happen to have periods comparable to those in Eq.~\eqref{eqn:Thf211}.

\section{Cluster and environment evolution}
\label{sec:evolution}

In Sec.~\ref{sec:df} we determined the impact of DF on the evolution of an individual star. To understand whether DF can be used to observationally constrain dense environments, in this section we study the consequences for the overall cluster of stars, as well as the cluster's effect on the environment. We focus for simplicity on a boson cloud in the $\ket{211}$ state, for which we have an explicit expression for the DF time, through Eq.~\eqref{eqn:Elost-cloud} and Fig.~\ref{fig:T_DF}.

\subsection{Distribution of orbital parameters}
\label{sec:simulation}

We perform a simplified simulation of the evolution of the S-cluster, as follows. We assume that the \emph{initial} density profile of the cluster is $\rho(r)=\rho\ped{s}(r\ped{s}/r)^\gamma$, as given in Eq.~\eqref{eqn:density-stellar-cusp}, that the cluster is isotropic, and that the potential is Keplerian, dominated by Sgr A*. The resulting phase-space distribution function of the stars, $f(E)$, is then only a function of the specific energy $E$. From $f(E)$ we determine the total mass $M\ped{s}(T)$ of stars with a period smaller than $T$. We then assume that all the stars in the cluster were born at the same time, with Kroupa's initial mass function (IMF) $\xi(M_*)$ \cite{Kroupa:2002ky}, and period $T<T\ped{max}$. We choose $T\ped{max}=\SI{1.5e5}{yr}$, as it corresponds to a semi-major axis equal to the influence radius of Sgr A*, $r_h=\SI{2.23}{pc}$.

From the isotropic phase space distribution $f(E)$ and the IMF $\xi(M_\star)$, we randomly sample the orbital parameters and the masses of a total number of stars equal to
\beq
N\ped{s}=\frac{M\ped{s}(T\ped{max})}{\braket{M_\star}}\,,
\eeq
where $\braket{M_\star}=\int_0^\infty\xi(M_\star)M_\star\dd M_\star$. For each star, we then calculate $T_\slab{df}$ and the scalar resonant relaxation timescale $T_\slab{srr}$ (see Appendix~\ref{sec:srr}). If $T_\slab{df}<T_\slab{srr}$, we evolve the orbit of the star under the effect of DF, assuming that its period decays according to Eq.~\eqref{eqn:T(t)_DF}. We remove a star from the cluster when $t>2T_\slab{df}(0)$, or when $t>T\ped{life}$, where $T\ped{life}$ is calculated as in Eq.~\eqref{eq:MSlife}.

\begin{figure*}
\centering
\includegraphics[]{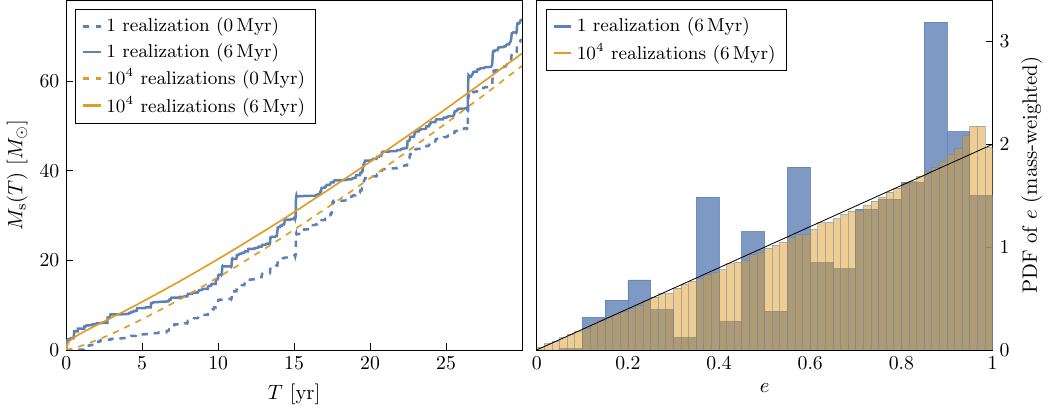}
\caption{\emph{Left panel:} cumulative mass of stars with period shorter than a given $T$. Dashed lines show the initial profile, calculated assuming an isotropic cluster with density $\rho(r)=\rho\ped{s}(r\ped{s}/r)^{\gamma\ped{s}}$. Solid lines show the evolved profile after $\SI{6}{Myr}$, assuming $M\ped{c}=0.05M$ and $\alpha=0.08$. \emph{Right panel:} mass-weighted distribution of eccentricities of stars with period less than $\SI{30}{yr}$. The thin black line shows the eccentricity distribution in an isotropic cluster, $p(e)\dd e=2e\dd e$.}
\label{fig:pop-evolution}
\end{figure*}

We show in the left panel of Fig.~\ref{fig:pop-evolution} the mass profile $M\ped{s}(T)$ of the cluster, both at $t=0$ and after evolving for $t=\SI{6}{Myr}$ (equal to the age of the Galactic Center disc \cite{Paumard:2006im}) in the presence of a $\ket{211}$ cloud with $M\ped{c}=0.05M$ and $\alpha=0.08$. Given the relatively high level of shot noise, we also show the result averaged over $10^4$ independent realizations of the star cluster. To make the calculation more manageable in this case, we used $T\ped{max}=\SI{3000}{yr}$, which we explicitly checked did not affect the results, thanks to $T_\slab{srr}$ being almost always shorter than $T_\slab{df}$ for stars with large period.

Because DF reduces the stars' periods, the cluster becomes denser at small stellar periods (such as $T<\SI{30}{yr}$). However, the effect is small, and is mostly washed out by the shot noise present in a single realization of the cluster. The stars most strongly affected by DF are the ones with high mass and on highly eccentric orbits. For this reason, we show in the right panel of Fig.~\ref{fig:pop-evolution} the mass-weighted distribution of eccentricities of the stars with $T<\SI{30}{yr}$. Once again, the dominant deviation from the isotropic distribution $p(e)\dd e=2e\dd e$ is caused by shot noise. The average over $10^4$ realizations reveals however that DF leads to a slight increase in the likelihood of a star having high eccentricity, roughly between $0.9<e<0.97$. This result may seem counterintuitive, as one might have expected a decrease in the number of stars on highly eccentric orbits, because of their efficient decay due to DF. However, for the same reason, among the very large number of stars with initial period $\SI{30}{yr}<T(0)<T\ped{max}$, those with $T_\slab{df}<T_\slab{srr}$ and extremely high eccentricity will also efficiently decay, and end up having period $T<\SI{30}{yr}$ while still retaining a large eccentricity. We also performed simulations where $T_\slab{srr}$ is neglected, as shown in Appendix~\ref{app:pop-evol-no-srr}. In this case, more stars decay through DF, which leads to a higher peak at large eccentricities.

Overall, while DF can severely alter the orbital evolution of an individual star, its impact on the global properties of the cluster is small. We attribute this to the fact that only a minority of stars have high enough mass and eccentricity for $T_\slab{df}$ to be short. Even when selectively looking at stars with the right properties, our simplified simulation shows that their number is significantly replenished by the decay of stars whose period was initially larger. Furthermore, it is likely that slight changes to the cluster's properties, such as its density profile or eccentricity distribution, suffer from degeneracies with the initial distribution function. We conclude that it is unlikely to observationally detect dense environments through DF, or constrain them with the existence of massive stars on eccentric orbits.

As a final remark, we note that our use of Kroupa's IMF should be a good approximation for S-stars, while the Galactic Center disc has been found to have a more top-heavy IMF \cite{2010ApJ...708..834B}. We re-ran the simulation with $\xi(M_\star)\propto M_\star^{-0.45}$, with a cutoff at $M_\star=100M_\odot$, and found that the shape of the (mass-weighted) eccentricity distribution robustly remains very similar to that in Fig.~\ref{fig:pop-evolution}. On the other hand, a more top-heavy IMF implies a smaller overall number of stars, and therefore increased shot noise.

\subsection{Impact of the star cluster on the environment}

After examining the effect of DF from the environment on the stellar population around Sgr A*, one may ask whether the cumulative energy lost by the decayed stars is sufficient to disrupt the environment. A star that goes from an initially large semi-major axis down to one comparable to the size of the environment, $r\ped{env}$, loses approximately an energy $\Delta U_\star\sim GM_\star M/r\ped{env}$. On the other hand, the environment's binding energy is $U\ped{env}\sim GM\ped{env}M/r\ped{env}$. This implies that the number of stars required to unbind the environment is
\beq
N_\star=\frac{U\ped{env}}{\Delta U_\star}\sim\frac{M\ped{env}}{M_\star}=\num{1.4e4}\biggl(\frac{M\ped{env}/M}{0.01}\biggr)\biggl(\frac{3\, M_\odot}{M_\star}\biggr).
\eeq
From the simulation described in Sec.~\ref{sec:simulation}, we find that, on average, about 9 stars completely decay over the course of the $\SI{6}{Myr}$, and that their average mass is around $2.7M_\odot$. We conclude that the stellar cluster does not have a large impact on the environment on $\si{Myr}$ timescales. If instead sustained decay of stars occurs over a much longer time interval $\Delta t$, the environment will lose a mass
\beq
\Delta M\ped{env}\sim10^{-4}M\biggl(\frac{\Delta t}{\SI{1}{Gyr}}\biggr)\,.
\eeq
Given the Galactic Center disc's age of $\SI{6}{Myr}$, it is unclear whether such a constant decay of stars can be sustained over $\si{Gyr}$ timescales. It would also be interesting to investigate if this continued mass loss could prevent the cloud's growth, and thus impact existing constraints on superradiance.

\section{Conclusions}
\label{sec:conclusions}

Accurate monitoring of S-star orbits around Sgr~A* remains one of the most powerful probes of the Milky Way's central potential, enabling precision tests of deviations from a Keplerian field and offering a rare window on BH environments and fundamental physics.

In this work, we derive new constraints on extended mass distributions around Sgr~A*, considering Plummer and power-law profiles and the axisymmetric densities expected for superradiant clouds. Using an orbit-averaged perturbative method, we compute the environment-induced apsidal precession and compare it directly with the measured precession of S2, avoiding numerical integration of the orbits.

For scalar clouds in the $\ket{211}$ state, our limits differ from Ref.~\cite{GRAVITY:2023cjt}: our results are consistent with zero at 1-$\sigma$, and we explore larger values of the boson-BH coupling $\alpha$. The cloud's quadrupolar field excludes cloud masses above $(0.1-1)\%\,M$ even when S2 lies outside its densest region. We also present the first constraints on $\ket{322}$ clouds. For a Plummer profile, we limit the total mass to $\lesssim 0.1\%\;M$ for scale radii comparable to S2's semi-major axis, and we bound the density of power-law profiles to within a factor of $\sim 30$ of the stellar-cluster expectation. Our limits complement and extend previous bounds ~\cite{GRAVITY:2023cjt,GRAVITY:2024tth}.

We further analyze secular orbital decay due to dynamical friction. We derive a general decay timescale formula, accurate up to $\mathcal{O}(1)$ geometric factors, and validated it with a detailed calculation for scalar clouds. Mass fractions of order $1\%\,M$ would lead to S-star decay within a few Myr. Yet, simulating the cluster evolution taking into account its dynamical timescales shows that inward migration of outer stars efficiently repopulates its inner region, making statistical detection of dynamical-friction signatures challenging. Furthermore, we show that orbital resonances from boson clouds have no impact on relevant timescales.

Alternative observables remain promising, e.g., spectroscopy along stellar trajectories could probe boson-photon couplings \cite{Bai:2025yxm}, and dense environments may impact tidal-disruption rates. Moreover, it is interesting to apply our framework to other well-motivated matter profiles around Sgr~A*, such as the dense cores (``solitons") predicted by ultralight dark matter~\cite{Bar:2019pnz}. The results for this analysis will be presented elsewhere~\cite{Caputo:2025ULDM}.

Future progress will come from improved orbital phase coverage and precision. With $\sim 30 \,\times$ smaller astrometric uncertainties, perturbations from other stars must be modeled explicitly. The most direct advance would be the discovery of stars with pericenters much closer than S2, e.g., the debated S62 and S4714 \cite{2020ApJ...889...61P,2020ApJ...899...50P}. Upcoming facilities---including GRAVITY+ at the VLTI, the next generation of extremely large telescopes (ELT, TMT, GMT), and high-resolution spectroscopy with JWST/NIRSpec---will substantially expand the accessible parameter space, offering a unique window on black-hole astrophysics and fundamental physics \cite{GRAVITYplus:Widmann2022,ELT:HARMONI2024,TMT:Stepp2012,GMT:McCarthy2018,JWST:NIRSpecJakobsen2022}.

\section*{Acknowledgments}

We thank Kfir Blum, Scott Tremaine and Matias Zaldarriaga for insightful conversations.
G.M.T. acknowledges support from the Sivian Fund at the Institute for Advanced Study. A.C. is supported by an ERC STG grant (``AstroDarkLS,'' Grant No.\ 101117510). A.C. acknowledges the hospitality of the Weizmann Institute of Science at different stages of this project. 
This project has received funding from the European Research Council
(ERC) under the European Union’s Horizon Europe research and innovation program (Grant Agreement No.\
101117510). Views and opinions expressed in this paper are
those of the authors only and do not necessarily reflect
those of the European Union. The European Union cannot be held responsible for them.

\newpage

\onecolumngrid

\appendix

\section{Calculation of apsidal precession}
\label{app:precession}

\subsection{Boson cloud}

The gravitational potential generated by the cloud's density profiles \eqref{eqn:density-cloud-211} and \eqref{eqn:density-cloud-322}, introduced schematically in Eq.~\eqref{eqn:potential-cloud-short}, can be written explicitly as
\begin{align}
\label{eqn:potential-cloud-211}
V_{211}(\vec r)={}&-\frac{M\ped{c}}{r}f_{211}^{(0)}\biggl(\frac{r}{r\ped{c}}\biggr)+3\frac{M\ped{c}r\ped{c}^2}{r^3}(3\cos^2\theta-1)f_{211}^{(2)}\biggl(\frac{r}{r\ped{c}}\biggr)\,,\\
\label{eqn:potential-cloud-322}
V_{322}(\vec r)={}&-\frac{M\ped{c}}{r}f_{322}^{(0)}\biggl(\frac{r}{r\ped{c}}\biggr)+18\frac{M\ped{c}r\ped{c}^2}{r^3}(3\cos^2\theta-1)f_{322}^{(2)}\biggl(\frac{r}{r\ped{c}}\biggr)-\frac{1215M\ped{c}r\ped{c}^4}{8r^5}(35\cos^4\theta-30\cos^2\theta+3)f_{322}^{(4)}\biggl(\frac{r}{r\ped{c}}\biggr)\,,
\end{align}
where
\begin{align}
f_{211}^{(0)}(x)&=1-e^{-x}\biggl(1+\frac{3x}4+\frac{x^2}4+\frac{x^3}{24}\biggr)\,,\\
f_{211}^{(2)}(x)&=1-e^{-x}\biggl(1+x+\frac{x^2}2+\frac{x^3}6+\frac{x^4}{24}+\frac{x^5}{144}\biggr)\,,\\
f_{322}^{(0)}(x)&=1-e^{-2x/3}\biggl(1+\frac59x+\frac{4x^2}{27}+\frac{2x^3}{81}+\frac{2x^4}{729}+\frac{2x^5}{10935}\biggr)\,,\\
f_{322}^{(2)}(x)&=1-e^{-2x/3}\biggl(1+\frac{2x}3+\frac{2x^2}9+\frac{4x^3}{81}+\frac{2x^4}{243}+\frac{11x^5}{10206}+\frac{5x^6}{45927}+\frac{x^7}{137781}\biggr)\,,\\
f_{322}^{(4)}(x)&=1-e^{-2x/3}\biggl(1+\frac{2x}3+\frac{2x^2}9+\frac{4x^3}{81}+\frac{2x^4}{243}+\frac{4x^5}{3645}+\frac{4x^6}{32805}+\frac{8x^7}{688905}+\frac{2x^8}{2066715}+\frac{2x^9}{31000725}\biggr)\,.
\end{align}
After setting $\theta=\pi/2$, we can insert Eq.~\eqref{eqn:potential-cloud-211} and Eq.~\eqref{eqn:potential-cloud-322} into \eqref{eqn:dotvarpi} to determine the apsidal precession rate. It is convenient to perform the differentiation with respect to $e$ under the integral sign, and write the result as
\beq
\dot\varpi=\frac{M\ped{c}\sqrt{1-e^2}}{2\pi r\ped{c}\sqrt{Ma}\,e}\int_0^{2\pi}F_{211/322}\biggl(\frac{r(E)}{r\ped{c}}\biggr)\cos E\dd E\,,
\label{eqn:dotvarpi-cloud}
\eeq
where $r(E)=a(1-e\cos E)$ and
\begin{align}
\label{eqn:F211}
F_{211}(x)&=\frac6{x^3}-e^{-x}\biggl(\frac6{x^3}+\frac6{x^2}+\frac3x+\frac54+\frac{x}2+\frac{3x^2}{16}+\frac{x^3}{16}\biggr)\,,\\
\label{eqn:F322}
F_{322}(x)&=\frac{36}{x^3}+\frac{3645}{2x^5}-e^{-2x/3}\biggl(\frac{3645}{2x^5}+\frac{1215}{x^4}+\frac{441}{x^3}+\frac{114}{x^2}+\frac{23}x+\frac{35}9+\frac{16x}{27}+\frac{7x^2}{81}+\frac{x^3}{81}+\frac{5x^4}{2916}+\frac{x^5}{4374}\biggr)\,.
\end{align}
The integral \eqref{eqn:dotvarpi-cloud} can be expressed in terms of Bessel functions for all terms in Eq.~\eqref{eqn:F211} and~\eqref{eqn:F322}, except those of the form $e^{-x}x^{-n}$ with $n>0$, in which case we could not find a simple closed form.

\subsection{Power-law density}

The gravitational potential generated by the power-law density profile in Eq.~\eqref{eqn:density-spike} is
\beq
V_0(r)=\frac{4\pi\rho_0r_0^{\gamma_0}r^{2-\gamma_0}}{(2-\gamma_0)(3-\gamma_0)}\,,
\eeq
whose orbit average is
\beq
\braket{V_0}=\frac{2\rho_0r_0^{\gamma_0}a^{2-\gamma_0}}{(2-\gamma_0)(3-\gamma_0)}\int_0^{2\pi}(1-e\cos E)^{3-\gamma_0}\dd E=\frac{4\pi\rho_0r_0^{\gamma_0}a^{2-\gamma_0}}{(2-\gamma_0)(3-\gamma_0)}\,{}_2F_1\biggl(\frac{\gamma_0-3}2,\frac{\gamma_0-2}2;1;e^2\biggr)\,.
\eeq
Taking the derivative of the hypergeometric function with respect to $e$ gives the result given in Eq.~\eqref{eqn:varpi-power-law}. This exact formula can be expanded in Taylor series for small eccentricity as
\beq
\dot\varpi=-T\rho_0\biggl(\frac{r_0}a\biggr)^{\gamma_0}\biggl(1-\frac{4+\gamma_0-\gamma_0^2}8\,e^2+\mathcal O(e^4)\biggr)\,.
\label{eqn:varpi-power-law-approx-small-2}
\eeq
For large $e$, one can instead approximate the hypergeometric function with its value at $e^2=1$, obtaining
\beq
\dot\varpi\approx-T\sqrt{1-e^2}\rho_0\biggl(\frac{r_0}a\biggr)^{\gamma_0}\frac{2\Gamma(5/2-\gamma_0)}{\Gamma((5-\gamma_0)/2)\Gamma((4-\gamma_0)/2)}\,.
\label{eqn:varpi-power-law}
\eeq
Equation \eqref{eqn:varpi-power-law} is extremely accurate when $\gamma_0$ is close to unity: for example, for the profile \eqref{eqn:density-stellar-cusp} with $\gamma\ped{s}=1.13$, the approximation is better than $3.1\%$ for any value of the eccentricity. Similar expressions can also be found in Refs.~\cite{Merritt2010, Ivanov2005, Polyachenko2007}.

\subsection{Plummer profile}

The gravitational potential generated by the Plummer profile in Eq.~\eqref{eqn:density-plummer} is
\beq
V\ped{pl}(r)=-\frac{M\ped{pl}}{\sqrt{\smash[b]{r\ped{pl}^2+r^2}}}\,,
\eeq
and the apsidal precession rate is
\beq
\dot\varpi=-\frac{M\ped{pl}\sqrt{1-e^2}}{2\pi r\ped{pl}\sqrt{Ma}\,e}\int_0^{2\pi}\frac{\cos E \dd E}{\bigl(1+r(E)^2/r\ped{pl}^2\bigr)^{3/2}}\,.
\eeq
The integral can be expanded for small eccentricities as
\beq
\dot\varpi=-\frac{3M\ped{pl}\sqrt{1-e^2}\,T}{4\pi r\ped{pl}^3(1+\Lambda)^{5/2}}\biggl(1+\frac{5\Lambda(4\Lambda-3)}{8(1+\Lambda)^2}e^2+\frac{35\Lambda^2(8\Lambda^2-20\Lambda+5)}{64(1+\Lambda)^4}e^4+\mathcal O(e^6)\biggr)\,,
\eeq
where $\Lambda=a^2/r\ped{pl}^2$. For $0.1<\Lambda<10$, this approximation is better than 20\% accurate for any value of the eccentricity.

\subsection{Generic spherically symmetric density}

We derive here a generalization of a well-known result for the apsidal precession in a general spherically symmetric density profile $\rho(r)$. We expand the potential $V(r)$ in a Taylor series as
\beq
V(r)\approx\sum_{n=0}^\infty\frac{(r-a)^n}{n!}\frac{\dd^nV}{\dd r^n}\bigg|_a\,.
\eeq
Defining $M(r)=\int_0^r4\pi r'^2\rho(r')\dd r'$, the first four derivatives are
\begin{align}
\frac{\dd V}{\dd r}&=\frac{M(r)}{r^2}\,,\\
\frac{\dd^2V}{\dd r^2}&=-\frac{2M(r)}{r^3}+4\pi\rho(r)\,,\\
\frac{\dd^3V}{\dd r^3}&=\frac{6M(r)}{r^4}-\frac{8\pi\rho(r)}{r}+4\pi\rho'(r)\,,\\
\frac{\dd^4V}{\dd r^4}&=-\frac{24M(r)}{r^5}+\frac{32\pi\rho(r)}{r^2}-\frac{8\pi\rho'(r)}r+4\pi\rho''(r)\,,
\end{align}
where $\rho'(r)$ and $\rho''(r)$ are the first and second derivatives of $\rho(r)$ with respect to $r$. The orbit-averaged values of the corresponding $(r-a)^n$ terms are
\begin{align}
\braket{(r-a)^2}=a\braket{(r-a)}&=a^2\frac{e^2}2\,,\\
\braket{(r-a)^4}=a\braket{(r-a)^3}&=a^4\frac{3e^4}8\,.
\end{align}
Plugging everything into Eq.~\eqref{eqn:dotvarpi}, we find
\beq
\dot\varpi\approx-T\biggl(\rho(a)+\biggl(\frac{\rho''(a)a^2}8+\frac{\rho'(a)a}4-\frac{\rho(a)}2\biggr)e^2+\mathcal O(e^4)\biggr)\,.
\label{eqn:dotvarpi-rho(r)-correction}
\eeq
This formula generalizes to second order in $e$ the well-known result $\dot\varpi\approx-T\rho(a)$ \cite{Tremaine:2004yu}, valid for quasi-circular orbits. When $\rho(r)$ is taken to be a power-law, Eq.~\eqref{eqn:dotvarpi-rho(r)-correction} reduces to Eq.~\eqref{eqn:varpi-power-law-approx-small-2}.

\vskip 20pt

\twocolumngrid

\section{Constraints on enclosed mass}
\label{app:constraints-enclosed-mass}

\begin{figure*}
\centering
\includegraphics[]{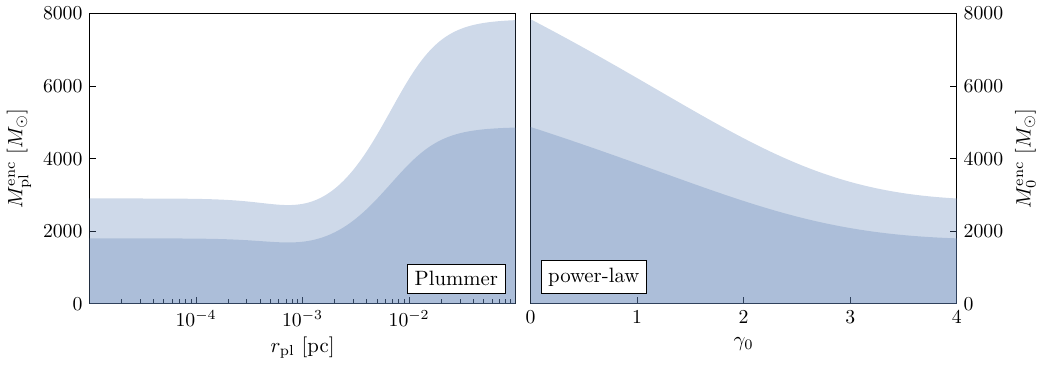}
\caption{1-$\sigma$ (dark blue) and 2-$\sigma$ (light blue) constraints on the environment mass enclosed between pericenter and apocenter of S2, as defined in Eq.~\eqref{eqn:enclosed-mass}, for the case of a Plummer (\emph{left panel}) and power-law (\emph{right panel}) density profiles.}
\label{fig:precession-bounds-plummer-spike-enclosed}
\end{figure*}

To allow for a more direct comparison with Ref.~\cite{GRAVITY:2024tth}, we translate here the constraints of Fig.~\ref{fig:precession-bounds-plummer-spike} into constraints on the environment's mass enclosed between pericenter and apocenter of S2, defined as
\beq
M\ap{enc}_{\mathrm{pl}/0}\equiv\int_{r\ped{p}}^{r\ped{a}}4\pi r^2\rho_{\mathrm{pl}/0}(r)\dd r\,,
\label{eqn:enclosed-mass}
\eeq
where $r\ped{a}=a(1+e)$. We show our result in Fig.~\ref{fig:precession-bounds-plummer-spike-enclosed}. This is in extremely good agreement with Fig.~B3 of Ref.~\cite{GRAVITY:2024tth}. The same paper also shows a multi-star analysis (see their Fig.~3), which provides slightly stronger constraints, with the 1-$\sigma$ bounds improving by about a factor of 2. As in Ref.~\cite{GRAVITY:2024tth}, our bound on $M\ap{enc}\ped{pl}$ is stronger at $r\ped{pl}\lesssim\SI{e-2}{pc}$. Note that the constraints on $M_0\ap{enc}$ and $M\ped{pl}\ap{enc}$ become identical when $\gamma_0=0$ and $r\ped{pl}\to\infty$, as both models reduce to a constant density environment.

\section{Star cluster timescales}
\label{sec:timescales}

Detecting or constraining environments through the apsidal precession or orbital decay of S-stars requires controlling for other astrophysical interactions. The current uncertainties on the apsidal precession are still too large for these corrections to matter; however, they will become important as astrometric data improve. On the other hand, the longer timescales of DF urge us to check whether any other astrophysical processes can change the star's orbit faster. In this appendix, we quantify the most important timescales of the stellar cluster. Useful reviews of the topic are available in Ref.~\cite{Alexander:2005jz} and in Fig.~1 of Ref.~\cite{2011MNRAS.412..187K}.

\subsection{Star lifetime and age}

The first timescale we consider is the expected lifetime of a star. Although this is not easy to determine precisely, a rough estimate will suffice for our purposes. The luminosity $L_\star$ of main sequence stars follows an approximately power-law relation with their mass, $L_\star\propto M_\star^{3.5}$~\citep{Kippenhahn2012}. As the total nuclear fuel is proportional to $M_\star$, we have $T\ped{life}\propto M_\star/L_\star$. Normalizing to the Sun's main sequence lifetime, we get
\beq
\label{eq:MSlife}
T\ped{life}=\SI{10}{Gyr}\biggl(\frac{M_\star}{M_\odot}\biggr)^{-2.5}=\begin{cases}
\SI{14}{Myr} & \text{S2},\\
\SI{32}{Myr} & \text{S29},\\
\SI{109}{Myr} & \text{S62},\\
\SI{1.8}{Gyr} & \text{S4714},\\
\end{cases}
\eeq
where we adopted the nominal values of the masses reported in Table~\ref{tab:stars}.

The age of the stars can also be estimated if spectroscopic measurements are available, for example by isochrone fitting using \texttt{MESA} Isochrones and Stellar Tracks (MIST) models~\cite{Dotter2016, Choi2016}. The age of S2 was estimated in Ref.~\cite{Habibi2017} as $T\ped{S2}=6.6^{+3.4}_{-4.7}\,\si{Myr}$, which is compatible with an age of $\SI{6}{Myr}$ of the Galactic Center disc \cite{Paumard:2006im} and also with the simple estimate \eqref{eq:MSlife}.

\subsection{Scalar resonant relaxation}
\label{sec:srr}

The orbit-averaged forces due to nearby stars induce a stochastic change in the magnitude $J$ of the orbital angular momentum, known as scalar resonant relaxation (SRR). The interaction can be imagined as if the mass of each star was spread out on a massive wire along its orbit \cite{RauchTremaine1996}. Assuming that all stars have the same mass $M\ped{eff}$, the typical specific torque applied by any such wire on a star with semi-major axis $a$ is $\tau_\star\sim M\ped{eff}/a$. When we add up the contribution of $N(a)$ randomly oriented wires, we get $\tau=\sqrt{N(a)}\beta(e)M\ped{eff}/a$, where $\beta(e)\approx0.25e$ is a dimensionless factor that quantifies the dependence on the eccentricity \cite{Hopman:2006qr,Gurkan:2007bj}, and the number $N(a)$ should be taken equal to the number of stars whose semi-major axis is less than $a$. These torques act coherently as long as the orientations of the orbits do not change. These orientations are scrambled in a time $T_{\dot\varpi}=2\pi/\abs{\dot\varpi}$ due to apsidal precession. The magnitude of the angular momentum therefore evolves as in a random walk, where each step has length $\Delta J=T_{\dot\varpi}\tau$, and the interval between consecutive steps is $T_{\dot\varpi}$. The time required to change $J=\sqrt{Ma(1-e^2)}$ by a factor of $\mathcal O(1)$ is thus
\beq
T_\slab{srr}\equiv\biggl(\frac{J}{\Delta J}\biggr)^2T_{\dot\varpi}=\frac{(1-e^2)M^2T^2}{4\pi^2\beta(e)^2M\ped{s}(a)M\ped{eff} T_{\dot\varpi}}\,,
\label{eqn:Tsrr}
\eeq
where we defined $M\ped{s}(a)=N(a)M\ped{eff}$. Given that the star's contributions add in quadrature, and that we reabsorbed one factor of $M\ped{eff}$ into $M\ped{s}(a)$, the effective mass appearing in Eq.~\eqref{eqn:Tsrr} should be taken to be $M\ped{eff}=\braket{M_*^2}/\braket{M_*}$. For Kroupa's IMF \cite{Kroupa:2002ky}, which we also adopt in Sec.~\ref{sec:evolution}, this is about $M\ped{eff}=3M_\odot$.

Note that the more commonly adopted definition of $T_\slab{srr}$ does not include the factor of $1-e^2$ in the numerator, which is equivalent to require that SRR changes the angular momentum by an amount equal to the angular momentum of a circular orbit with radius $a$. For the purposes of environment-induced apsidal precession and DF, however, it is crucial that the star's pericenter remains within the densest region of the environment. Our prescription takes this aspect into account, as on very eccentric orbits the angular momentum is directly related to $r\ped{p}$ via $J\approx\sqrt{2Mr\ped{p}}$.

If the precession timescale is $T_{\dot\varpi}=2\pi/\dot\varpi_\slab{gr}$, as given in Eq.~\eqref{eqn:gr-precession}, and we use $M\ped{s}(a)$ as computed from Eq.~\eqref{eqn:density-stellar-cusp} assuming the cluster is isotropic, we get
\beq
\label{eq:TRR}
T_\slab{srr}=\begin{cases}
\SI{131}{Myr} & \text{S2},\\
\SI{23}{Myr} & \text{S29},\\
\SI{170}{Myr} & \text{S62},\\
\SI{139}{Myr} & \text{S4714}.\\
\end{cases}
\eeq
For these S-stars, SRR is slower than DF, unless $M\ped{env}\lesssim10^{-3}M$ (or $M\ped{env}\lesssim10^{-2}M$ for S29).

For stars with a much larger period, the conclusion is different. Stars that experience strong DF necessarily have $r\ped{env}\sim r\ped{p}=a(1-e)$, therefore $1-e$ scales inversely with $a$. Under this assumption, we get that $T_\slab{srr}$ decreases as $a^{\gamma\ped{s}-5/2}$, while $T_\slab{df}^{(0)}$ increases as $a^{1/2}$. More precisely, we find that $T_\slab{srr}<T_\slab{df}^{(0)}$ for
\beq
a\gtrsim\SI{0.09}{pc}\biggl(\frac{M_\star}{10M_\odot}\;\frac{500M}{r\ped{env}}\;\frac{M\ped{env}}{0.01M}\biggr)^{0.54}\,,
\label{eqn:a_Tsrr_Tdf}
\eeq
corresponding to a period of $T\gtrsim\SI{1250}{yr}$ for these reference values of $M_\star$, $r\ped{env}$ and $M\ped{env}$.

Finally, we note that the SRR time is affected by the environment-induced apsidal precession through $T_{\dot\varpi}$. As one can see from Fig.~\ref{fig:precession-per-orbit}, the peak magnitude of $\delta\varpi$ depends on $M\ped{env}$ but not much on $r\ped{p}$. At fixed $r\ped{p}\sim r\ped{env}$, both the general relativistic and environmental contribution to $T_{\dot\varpi}$ scale linearly with $T$. Given the observational constraints from S2, a safe assumption is then that for stars with similar pericenter the relativistic precession always dominates, in which case the estimates \eqref{eq:TRR} and \eqref{eqn:a_Tsrr_Tdf} are justified. For stars with much larger $r\ped{p}$, instead, the retrograde precession due to the star cluster will eventually dominate.\footnote{If a supermassive BH has a very dense environment, the point where the prograde and retrograde contributions to the precession cancel might be shifted. This is particularly relevant for extreme mass ratio inspirals, where relativistic precession is believed to give rise to the so-called ``Schwarzschild barrier''~\cite{Merritt:2011ve}, which limits the effectiveness of stellar torques, and thus suppresses the rate at which stars are captured by the central black hole.}

\subsection{Vector resonant relaxation}

Similar to SRR, the orientation of the angular momentum is also affected by the orbit-averaged forces of nearby stars, a process known as vector resonant relaxation (VRR). The VRR timescale is~\cite{2011MNRAS.412..187K, Eilon:2008ht}
\begin{equation}
T_\slab{vrr}=\frac{MT}{\beta_v^2\sqrt{M\ped{s}(a)M\ped{eff}}} \,,
\end{equation}
where $\beta_v\approx1.83$. Its value for the S-stars is
\beq
\label{eq:T_VRR}
T_\slab{vrr}=\begin{cases}
\SI{1.7}{Myr} & \text{S2},\\
\SI{3.3}{Myr} & \text{S29},\\
\SI{1.4}{Myr} & \text{S62},\\
\SI{1.5}{Myr} & \text{S4714}.\\
\end{cases}
\eeq
VRR is much faster than SRR, because the coherence of its torques is broken by nodal precession, rather than by the much faster apsidal precession. However, VRR does not change the pericenter distance of the stars, and is thus not particularly relevant for DF. In cases of non-spherical environments, such as boson clouds, VRR can lead to a modulation of $T_\slab{df}$ as the orbital plane changes over time.

\subsection{Physical collisions}

In the inner parts of the cluster, physical collisions between stars could become more likely. For a cluster with density $\rho$, composed of stars with equal mass $M_\star$, radius $R_\star$ and velocity dispersion $\sigma$, the collision timescale is~\cite{Binney:2008}
\begin{equation}
T\ped{coll}=\frac{M_\star}{16\sqrt\pi\rho\sigma R_{\star}^2 (1 + \Theta)}\,,\qquad\Theta=\frac{M_\star}{2\sigma^2R_\star}\,.
\end{equation}
Assuming $M_\star=M_\odot$, $R_\star=R_\odot$, and a velocity dispersion $\sigma(r)\approx\SI{280}{km/s}\sqrt{\SI{0.1}{pc}/r}$ \cite{2011MNRAS.412..187K}, the term $\Theta$ is negligible for small $r$, and we obtain
\beq
T\ped{coll}=\SI{69}{Myr}\,\biggl(\frac{r}{\SI{e-3}{pc}}\biggr)^{1.63}\,.
\eeq
The collision timescale is generally longer than the other timescales analyzed, although it might be shorter than $T\ped{life}$ for stars with low enough mass.

\subsection{Two-body relaxation}

The two-body relaxation time for S-cluster stars is given by~\cite{Binney:2008}
\beq
T\ped{relax}=\frac{0.34\sigma^3}{\rho M\ped{eff}\log\Lambda}=\SI{25}{Gyr}\biggl(\frac{r}{\SI{e-3}{pc}}\biggr)^{-0.37}\,.
\eeq
This timescale is extremely long and therefore of limited relevance to the phenomenology discussed in this paper.

\section{Cluster evolution neglecting scalar resonant relaxation}
\label{app:pop-evol-no-srr}

\begin{figure*}
\centering
\includegraphics[]{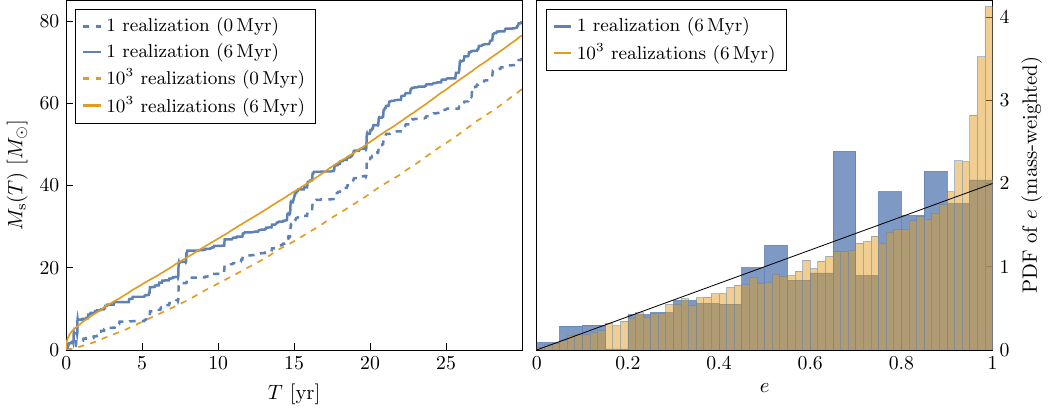}
\caption{Same as Fig.~\ref{fig:pop-evolution}, but neglecting the SRR timescale.}
\label{fig:pop-evolution-no-srr}
\end{figure*}

We show in Fig.~\ref{fig:pop-evolution-no-srr} the results of a simulation like the one described in Sec.~\ref{sec:simulation}, but allowing the stars with $T_\slab{srr}<T_\slab{df}$ to decay. Although a single realization of the evolved cluster is still largely dominated by shot noise, the average over $10^3$ realizations shows an increase in the density higher than that observed in Fig.~\ref{fig:pop-evolution}, and a much higher increase in the number of stars having high eccentricities. This is because, when $T\ped{max}$ increases, the total number of stars increases faster than the probability that they have $r\ped{p}\lesssim r\ped{env}$ decreases. Most of the stars experiencing DF have therefore very large periods, if SRR is neglected. Note that, for this reason, we could not improve the computational cost by reducing $T\ped{max}$ when performing the average over $10^3$ realizations in this case.

%\clearpage
\bibliography{main}

\end{document}